# Green production as a factor of survival for innovative startups. Evidence from Italy

Riccardo Gianluigi Serio[1], Maria Michela Dickson[2], Diego Giuliani[2] and Giuseppe Espa[2]

[1] – Faculty of Law, University of Trento
[2] – Department of Economics and Management, University of Trento

## Abstract

Many studies have analyzed empirically the determinants of survival for innovative startup companies using data about the characteristics of entrepreneurs and management or focusing on firm- and industry-specific variables. However, no attempts have been made so far to assess the role of the environmental sustainability of the production process. Based on data describing the characteristics of the Italian innovative startups in the period 2009-2018, this article studies the differences in survival between green and non-green companies. We show that, while controlling for other confounding factors, startups characterized by a green production process tend to survive longer than their counterparts. In particular, we estimate that a green innovative startup is more than twice as likely to survive than a non-green one. This evidence may support the idea that environment sustainability can help economic development.

# 1. Introduction

Innovation has always been a distinguishing activity of humanity. From the wheel to the invention of the world wide web, it constitutes an indissoluble common thread, inherent in human nature itself. Innovation is also crucial for economic growth and sustainability. Following the growing pressure brought about by international competition, it is now a duty for companies and modern nations to innovate. An important component of the engine of innovation is to be found in *startup* companies. The term startup appears for the first time on Forbes in 1976 (Forbes, 1976) to indicate a new type of company and not an embryonic phase of the life of incumbent companies. Since then, the attention on the topic has been growing due to the occurrence of positive association among new business initiatives, innovation rate and economic growth (see, among others, Kirchhoff et al., 2007; Baptista et al., 2008; Bygrave et al., 2003; Colombelli et al., 2016)

Startups may stimulate economies by promoting innovation (Audretsch, 1995; Reynolds, 1997). The most innovative ones often report better performances (Vivarelli and Audretsch, 1998), contribute positively to the generation of new jobs and to the development of new sectors (Acs and Audretsch, 1987; Shearman and Burrell, 1988), and, more in general, fuel an overall improvement of the welfare system (Birch, 1979, 1987; Phillips and Kirchhoff, 1989; Rickne and Jacobsson, 1999).

Despite their contribution to innovation and economic development, startups often struggle to survive for a long time in the market, because of the higher difficulties they face, especially at the beginning of their business activity. The greatest problem that newborn companies deal with concern the so-called liability of newness (Stinchcombe, 1965), which is usually associated with the scarcity of resources that new entries have access to start and develop the business. Indeed, some balance sheet indicators, such as a consistent liquidity, very low leverage ratio and the ability to make profits, have been shown to be important predictors of success or failure of a startup, especially in the preliminary stages of its life (Wiklund et al., 2010). Other features that may affect the positive result of a newborn business activity concern the characteristics of the entrepreneur, e.g. his/her previous work experiences in the business world (Lazear, 2004; Dahl and Reichstein, 2007) or his/her ability to cooperate (Eisenhardt, Schoonhoven, 1990). As a result, newborn companies, and hence startups, may experience high failure rates (Shapero and Giglierano, 1982), especially those that reproduce existing products (Finaldi Russo et al., 2016).

Recently, innovation has been increasingly dedicated to problem of using and searching for alternative energy sources to mitigate the impact of human activities on Earth, according to the so-called *green economy*. Indeed, pursuing a sustainable development is more and more a global issue (Finnegan et al., 2018; Garbasso, 2014; Crespi et al., 2015).

Thus, when the will to innovate meets the need to pursue sustainability, innovative startups play a crucial role (Iazzolino et al., 2019). Given that, it is reasonable to expect some form of attention and protection by the policy maker aimed at promoting the development of new business initiatives focused on technological sustainable innovation. Some studies have reported evidences that seem to validate this assumption (e.g. Söderblom and Samuelsson, 2014). The goal of policy makers is to intensify technology transfer and market competition to speed up the evolution of the industrial network and hence to increase the production and the employment (Autio and Parhankangas, 1998; Ejermo and Xiao, 2014; Storey and Tether, 1998). In particular, it has been shown that sustainability-oriented technologies offer the opportunity to restore competitiveness in western saturated mature economies (Mazzanti and Zoboli, 2009; Costantini et al., 2013; Gilli et al. 2014). Indeed, in the last years, the green economy has been one of the best responses to the economic crisis. Not surprisingly, also startups have been interested by green consciousness, given the possibility to receive additional benefits if adopting environmental ethics.

However, while it is true that companies devoted to a sustainable production have recently gain an increase in growth, this not necessarily goes along with the probability to survive in the market, which should be a central topic for startups. Although several contributions in literature addressed the issue of identifying the factors of survival (e.g. Arbia et al., 2017) and development of young innovative companies (e.g. Giraudo et al., 2019), few studies have focused on investigating the link between sustainable development and neo-entrepreneurial activity (Schick, et al., 2002).

This paper aims at filling this gap in the literature by assessing whether and how the risk of market exit that innovative startups face is affected by the environmental sustainability of their production process. In particular, exploiting data about the population of Italian innovative startups in the period 2009-2018 and by means of the Cox proportional-hazard model, we verify that *green* startups have a relatively higher survival performance compared to the *non-green* ones, while controlling for other structural factors that influence firm survival.

The paper is structured as follow. Section 2 briefly describe the current Italian legislative framework of innovative startups, that is the Italian Startup Act. In Section 3, the statistical framework is presented. In Section 4 the dataset about Italian innovative startups on which the analyses are conducted is presented, and interesting insights are brought to light. Section 5 concludes.

## 2. The Italian legislative framework of innovative startups: the "Startup Act"

At the end of 2012 the Italian government decided to intervene to improve the context for the birth and growth of newborn companies through the so-called *Italian Startup Act* (Law no. 221/2012), which introduces the definition of a new innovative company in the Italian legal system, namely the *innovative startup* (hereinafter ISU). Requirements to be considered ISU are related to the nationality (being registered in Italy or in another EU country but with a production branch in Italy), age (aged less than 5 years) and core business (which must be centered on research, development, production, marketing of innovative products with high technological value) of the company. In addition, ISUs must satisfy at least two of three additional requirements, such as: expenses in R&D and innovation must be at least 15% of either its annual costs or its turnover; employs highly qualified personnel, such as at least one third of PhD holders and students, or researchers, or at least two third of M.Sc. graduates; be the owner, depositary or licensee of a registered patent, or the owner of a registered software (Ministry of Economic Development, 2019). Newborn companies respecting these requirements may be registered in the special section for innovative startups of the Italian Business Register. The aim of the Italian legislator was clearly the creation of an environment for the development of new entrepreneurial ideas with a highly innovative character. So that, the possibility to be defined ISUs was contemplated also for companies that comply with the requirements for a period **(i)** not exceeding four years, if the company was born up to 2 years before entry into force, **(ii)** not exceeding three years, if established between 2009 and 2010, and **(iii)** up to two years, if registered in the Italian Business Register between 2008 and 2009. Companies falling into definition of ISUs may enjoy a substantial number of concessions, going from the purely bureaucratic sphere to tax, financing and governance grants. Examples of the first are the cut of many red tape rules and exemption to pay annual fees and duty stamps, while examples of the second are tax incentives for equity investors, an easier compensation of VAT credits, the extension of terms to cover systematic losses, a flexible corporate managing, tailor-made labor laws, remuneration of employees and consultants through stock options and work for equity (not included in taxable income), not compulsory operationality tests to verify the inactivity status, facilitated and speed-up bankruptcy procedures, and many others (Finaldi Russo et al., 2016; Ministry of Economic Development, 2019).

The Italian regulation on ISUs provides a particular kind of innovative startup defined as "*high technological value companies in energy related fields*". These companies are intended as *green* startups (in contrast with *non-green* startups), which shall establish green oriented activities, regardless of their specific sectors of activity. Business literature and empirical evidences show that nowadays sustainability and innovation go hand in hand and feed into themselves. Adopting a green approach means, for newborn companies, to transform initial difficulties into opportunities. Although Italy has been lagging in the green transition, the last GreenItaly report (Unioncamere and Fondazione Symbola, 2019) has pointed out that the 31.2% of the entire non-agricultural entrepreneurship has invested in the period 2015-2018, or plan to invest by the end of 2019, in green products and technologies in order to reduce the environmental impact, save energy and curbing $CO_2$ emissions. Therefore, we argue that the innovative startups belonging to this class can be properly considered as *green* as opposed to the *non-green* ones. This categorization can then help in assessing whether *greenness* positively affects the survival performance of innovative startups.

## 3. Empirical methodology: survival analysis

The proper empirical methodology to assess the determinants of survival time of a company, that is the time occurring between the entry of a company into the market and its exit from the market, is the so-called survival (or duration) analysis. Unlike the more traditional regression modelling approaches, such as the logistic regression, survival analysis can specifically deal with the inevitable occurrence of censoring, that is the presence of truncated observations due to the fact that the actual survival time of a company can be longer than its observed follow-up time. In particular, in order to study the relationship between company survival performance and *greenness* in production, we employ Kaplan-Meier curves and the Cox proportional hazards regression model.

*3.1 Descriptive survival analysis: the Kaplan-Meier curves*

Following the approach by Kaplan and Meier (1958), it is possible to estimate the company survival probability non-parametrically using the observed survival times, both censored and uncensored, of each company.

Let consider that $k$ companies cease to operate during the time interval under observation at distinct points in time $t_1 < t_2 < t_3 < \cdots < t_k$. Assuming that exit of companies from the market occur independently of one another, the probabilities of surviving from a point in time to the successive one can be multiplied together to give the cumulative survival probability. In other words, the probability that a company is still on the market at $t_j$, say $S(t_j)$, can be calculated from $S(t_{j-1})$ as follows:

$$S(t_j) = S(t_{j-1})\left(1 - \frac{d_j}{n_j}\right),$$

where $n_j$ is the number of companies still being on the market just before $t_j$ and $d_j$ represents the number of companies that exit from market at $t_j$. Obviously, since $t_0 = 0$, then $S(0) = 1$. The value of $S(t)$ is necessarily constant between successive points in times thus implying that the estimated probability is a step function that varies only at the point in time of each exit. The Kaplan-Meier (hereinafter KM) survival curve is the plot of $S(t)$ against $t$ and may provide a useful summary of the survival performance of companies.

In addition, the comparison between the KM curves of different subgroups of companies, such as the groups of green and non-green startups, allows to identify the presence of factors affecting the survival. Indeed, it is possible to test the statistical significance of the difference between the survival curves of different groups through the log-rank test (Peto et al., 1977). This test is based on the computation, for each group, of the expected number of companies that cease to operate at each point in time, since the previous one, under the null hypothesis of no difference between groups. For each $i$-th group, the sum of these values across all points in time provides the total expected number of companies' exits, say $E_i$. The log-rank test summarizes the discrepancies between the observed number of companies' exits in each group, say $O_i$, and $E_i$ by means of the following test statistic,

$$\chi^2 = \sum_{i=1}^{g} \frac{(O_i - E_i)^2}{E_i}.$$

Under the null hypothesis of no difference between the survival curves of the groups, the $\chi^2$ test statistic follows a Chi-square distribution with $(g-1)$ degrees of freedom, where $g$ is the number of groups.

*3.2 Survival regression modelling: the Cox proportional hazards model*

Regression methods for survival time data attempt to model the relationship between one or more regressors and the so-called hazard function $\lambda_i(t)$, which in this empirical context denotes the instantaneous exit rate for company $i$ surviving to time $t$. Consequently, $\lambda_i(t)dt$ gives the probability of company $i$ to exit from market at time $t$, given that it survived until time $t$. Unlike other parametric survival regression models, which require to specify a functional form for $\lambda_i(t)$, the semi-parametric *Cox proportional-hazards model* (Cox, 1972) does not require to make any distributional assumption.

Under this model, the hazard function for company $i$ varies according to time $t$ and $k$ regressors $(x_1, x_2, \ldots, x_k)$ as follows:

$$\lambda_i(t) = \lambda_0(t)\exp(\beta_1 x_{1i} + \beta_2 x_{2i} + \cdots + \beta_k x_{ki})$$

where $\lambda_0(t)$ represents the baseline hazard and $\beta_1, \beta_2, \ldots, \beta_k$ are unknown parameters that need to be estimated. In this formulation there is no need to specify the functional form of $\lambda_0(t)$ since it is assumed to be common among all companies. Indeed, the ratio between the hazards of any two generic companies $i$ and $l$ is

$$\frac{\lambda_i(t)}{\lambda_l(t)} = \frac{\lambda_0(t)\exp(\beta_1 x_{1i} + \cdots + \beta_k x_{ki})}{\lambda_0(t)\exp(\beta_1 x_{1l} + \cdots + \beta_k x_{kl})} = \exp[\beta_1(x_{1i} - x_{1l}) + \cdots + \beta_k(x_{ki} - x_{kl})]$$

and, therefore, it does not depend on neither $t$ nor $\lambda_0(\cdot)$.

Although the specified model does not make any assumption about the data generating process, it however needs that the hazards are proportional. The holding of the proportional-hazards assumption can be verified with the Grambsch-Therneau P.H. test (Grambsch and Therneau, 1994). The estimation of the model parameters and the associated significance tests can be achieved using the partial likelihood technique (Cox, 1975).

## 4. Results

Data used to perform the analysis proposed in the present paper concern Italian ISUs and cover the period 2009-2018, which corresponds to the period of the Startup Act. In order to avoid spurious results, ISUs in agricultural sector have been excluded from the analyses because they are subject to a different legislation about business failure. The first five years of time span have been grouped in one category, due to the small number of units in each year. The total number of startup companies at the end of the period was equal to 9,453. Table 1 reports the total number and the percentage on the total of innovative startup companies, distributed by Italian macro-area. At the end of the time span, more than half of the startups (55.1%) are located in northern Italy (23.3% in North-East and 31.8% in North-West), while 20.5% are located in central Italy and only under a quarter (24.4%) in southern Italy. Compared to the beginning of the period, the growth in the number of startups has involved both northern and southern areas of the country, with only central Italy to have suffered a considerable decrease.

**Table 1**: Registered ISUs in Italy by macro-area: numbers (n) and percentages on the total (%) at the end of each year.

| Macro-area | Until to 12.31.2013 | | 12.31.2014 | | 12.31.2015 | | 12.31.2016 | | 12.31.2017 | | 12.31.2018 | |
|---|---|---|---|---|---|---|---|---|---|---|---|---|
| | *n* | *%* | *n* | *%* | *n* | *%* | *n* | *%* | *n* | *%* | *n* | *%* |
| North-East | 113 | 20.1 | 388 | 20.9 | 786 | 22.1 | 1,250 | 22.9 | 1,883 | 23.6 | 2,207 | 23.3 |
| North-West | 174 | 31.0 | 558 | 30.0 | 1,081 | 30.4 | 1,707 | 31.2 | 2,495 | 31.3 | 3,004 | 31.8 |
| Center | 141 | 25.1 | 391 | 21.0 | 772 | 21.7 | 1,148 | 21.0 | 1,636 | 20.5 | 1,940 | 20.5 |
| South | 133 | 23.7 | 522 | 28.1 | 915 | 25.7 | 1,358 | 24.9 | 1,959 | 24.6 | 2,302 | 24.4 |
| Total | 561 | 100 | 1,859 | 100 | 3,554 | 100 | 5,463 | 100 | 7,973 | 100 | 9,453 | 100 |

Regarding the economic sector of activity (according to the NACE classification), most of ISUs belong to the service sector, which is the 76.3% of the total startup companies at the end of the reference period. The manufacturing sector, with its 18.4% on the total, constitutes a not negligible share, while startups operating in tourism and trade form together little more than 5% (Table 2).

**Table 2**: Registered ISUs in Italy by macro-sector of activity: numbers (*n*) and percentages on the total (%) at the end of each year.

| Sector | Until to 12.31.2013 | | 12.31.2014 | | 12.31.2015 | | 12.31.2016 | | 12.31.2017 | | 12.31.2018 | |
|---|---|---|---|---|---|---|---|---|---|---|---|---|
| | *n* | *%* | *n* | *%* | *n* | *%* | *n* | *%* | *n* | *%* | *n* | *%* |
| Manufacturing | 95 | 16.9 | 326 | 17.5 | 631 | 17.8 | 1,008 | 18.5 | 1,474 | 18.5 | 1,736 | 18.4 |
| Services | 434 | 77.4 | 1,437 | 77.3 | 2,721 | 76.6 | 4,140 | 75.8 | 6,053 | 75.9 | 7,208 | 76.3 |
| Tourism | 6 | 1.1 | 13 | 0.7 | 31 | 0.9 | 49 | 0.9 | 81 | 1.0 | 92 | 1.0 |
| Trade | 26 | 4.6 | 83 | 4.5 | 171 | 4.8 | 266 | 4.9 | 365 | 4.6 | 4,17 | 4.4 |
| Total | 561 | 100 | 1,859 | 100 | 3,554 | 100 | 5,463 | 100 | 7,973 | 100 | 9,453 | 100 |

The first research question that we aim to address concerns to verify if green innovative startup companies have a higher or lower survival performance compared to the non-green innovative startups. In order to address this question, we have computed the KM survival curves and implemented the Cox Proportional Hazard model. In using the described dataset, we had to deal with several missing values in some variables. Therefore, the number of observations varies among the different analyses.

Panel (a) of Figure 1 shows the KM curves for 9,453 ISUs in Italy in the 2009-2018 period. At the estimated survival probability after 8 years of activity for is nearly 57.5%, indicating that more than half of the considered companies is still in the market at the end of the observational period.

**Figure 1**: Kaplan–Meier estimates of survival probability of 9,453 ISUs in Italy for the period 2009–2018. On the *x*-axis is reported the time (in number of months) from when a company enters the market, defined as first registration in the Italian business register. On the *y*-axis is reported the cumulative survival probability. Panel (a) shows the Kaplan-Meier curve for all considered startups, while panel (b) shows Kaplan-Meier curves for green and non-green ISUs (Logrank test $\chi^2 = 13.7$, *p*-value=0.000).

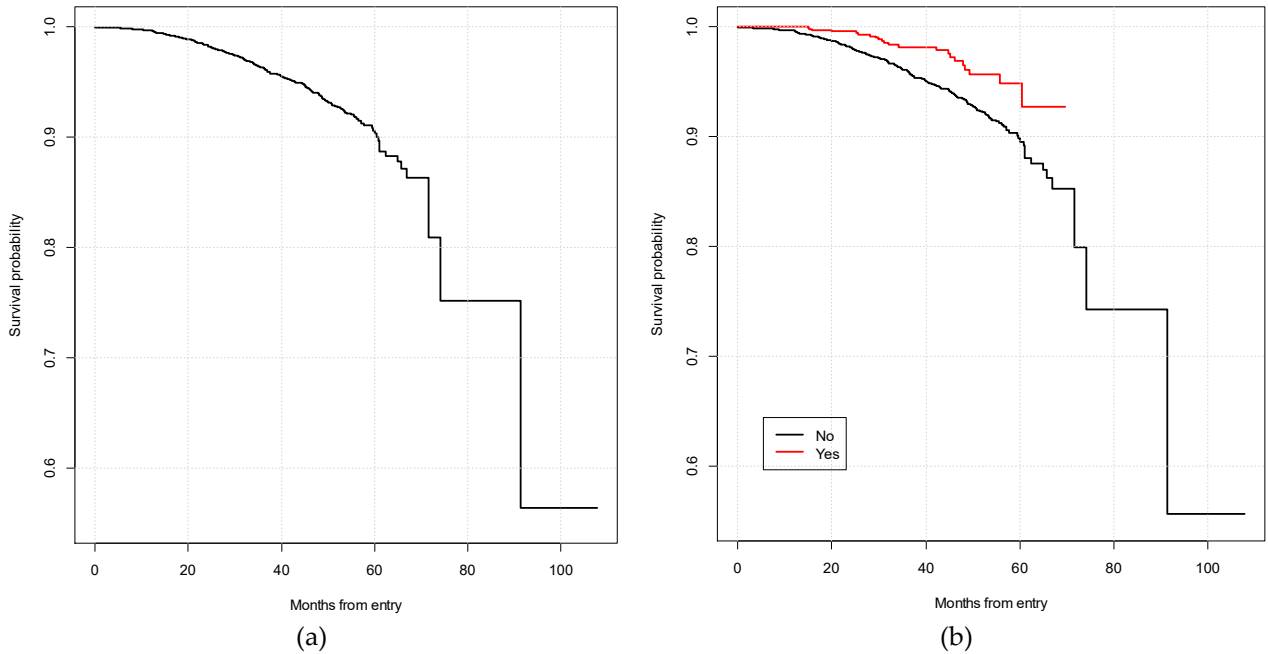

(a)          (b)

Focusing the attention on comparing green and non-green innovative startups, Figure 1(b) shows that the survival performance of the former is very high, so that over 92% of the companies observed do not experience the exit from the market. The conducted analysis highlights a clear distinction in the performance of green startup companies compared to non-green. In particular, the formers have a substantially better survival performance with respect to the latter ones. With a *p*-value approximately equal to zero, the Logrank test shows also that the difference between their survival curves is indeed statistically significant.

This difference, however, could be at least partially due to the particular regulation of the innovative startups instead of structural differences between green and non-green companies. In fact, since the introduction of regulation 221/2012 and especially from the introduction of regulation 221/2015 to promote the green economy, considerable benefits are granted to stimulate the run-up to sustainability. Innovative startup companies that promote the achievement of this aim may benefit of substantial reduction in costs and tax charges, allowing them to not incur in liquidity crisis, unsustainable leverage and difficulty in finding funding sources, which are only few examples of decisive causes of the premature exit from the market (Stinchcombe, 1965; Baum, 1996, Hannan and Freeman, 1984; Wiklund et al., 2010).

This consideration is supported by the results provided by Figure 2, which shows the impact of two economic and financial performance measures on the survival of ISUs, that is the return on assets (ROA) indicator and the debt-to-equity ratio (D/E). The choice to use these measures to capture the effect of the financial characteristics of companies is twofold. Firstly, other items in the financial statements had a large number of missing; in fact, it is possible to note that the total number of units is lower than that previously used, and startups considered are all born before 2017. Secondly, this information will be used as control variables in the following analyses. For both the ROA and D/E indicators, firms are grouped into quartiles to simply illustrate the differences between companies with higher and lower values (Figure 2).

**Figure 2:** Panel (a): Kaplan–Meier survival curves of 5,327 ISUs in Italy, born before 2017, analyzed by ROA for the period 2009–2018 (Logrank test $\chi^2 = 160$, *p*-value=0.000). Panel (b): Kaplan–Meier survival curves of 5,014 ISUs in Italy, born before 2017, analyzed by debt to equity ratio for the period 2009–2018 (Logrank test $\chi^2 = 41.7$, *p*-value=0.000).

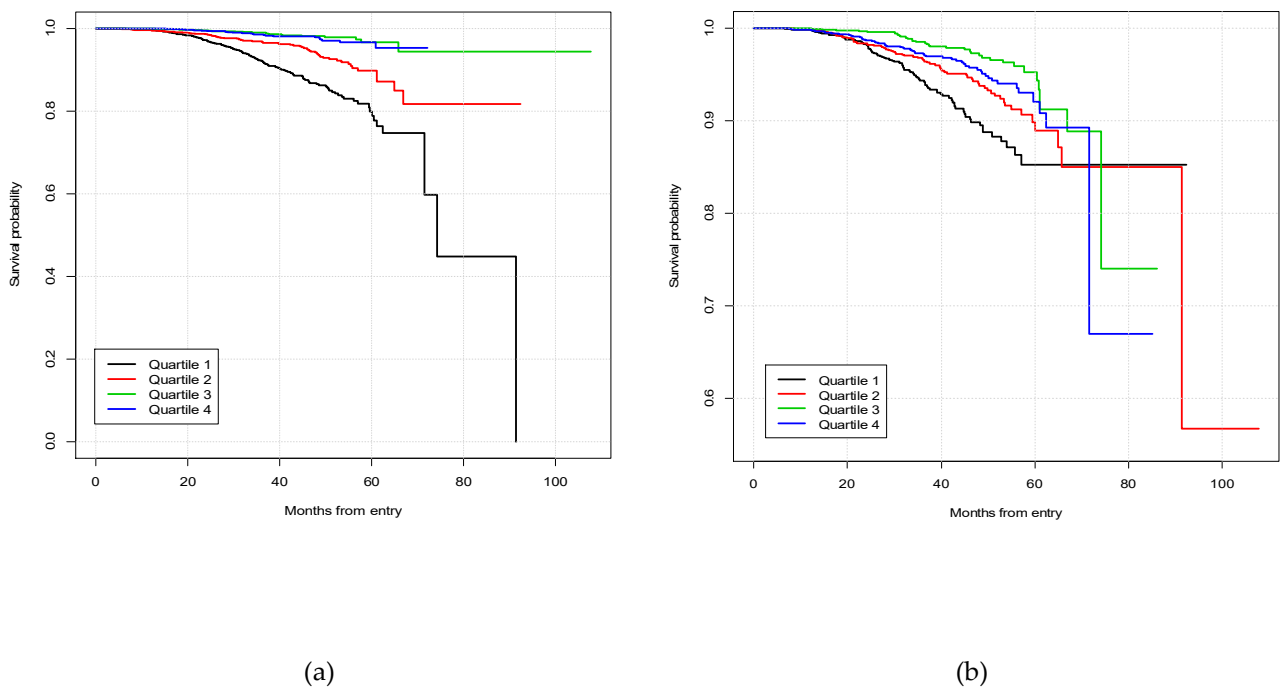

(a)  (b)

Results of the analysis show that companies with a high ability to remunerate assets, have a better response in terms of survival probability. Companies with a ROA value located in the third and fourth quartiles of the distribution have a survival probability that is never less than 90% in the considered period. Concerning the D/E, findings are in line with the literature and highlight how a lower pressure of debt on capital favors the chances of survival. Low levels of debt can increase survival probability in at least two ways. On one hand, this makes the future cash flows of a company more secure, as they will not be absorbed as a priority by the debt holders but can instead be used for normal business operations or further investments (Panno, 2003). On the other hand, companies with a low leverage ratio, can look at new debt as a valid path to be taken to meet future needs (Wiklund et al., 2010).

Therefore, in order to assess the effect of greenness in production on the survival performance of innovative startups while controlling for their financial conditions, as measured by ROA and D/E indicators, we also fitted a Cox proportional-hazard model. We estimated two models with three explanatory variables (Model 1 and Model 2) and we introduced the two control variables, namely ROA and D/E, in Model 3. Results of the estimation are reported in Table 3.

Model 1 has been estimated on the entire dataset covering the entire period, while Model 2 and Model 3 have been fitted only on startup companies born before 2017, due to the high number of missing elements present in the control variables for the last two years.

**Table 3**: Results from the estimation of Cox proportional hazard models. Coefficients and Standard Errors (in parentheses) are reported for each explanatory variable used in the model.

|  | All companies | Companies born before 2017 | |
|---|---|---|---|
|  | Model 1 | Model 2 | Model 3 |
| EnergyRelated[YES] | -0.79 (0.22)*** | -0.72 (0.23)*** | -0.63 (0.24)*** |
| Industry[Services] | 0.35 (0.17)** | 0.32 (0.18)* | 0.30 (0.19) |
| Industry[Tourism] | 1.17 (0.41)*** | 1.17 (0.44)*** | 0.88 (0.53)* |
| Industry[Trade] | 0.49 (0.28)* | 0.51 (0.29)* | 0.43 (0.30) |
| ROA[quartile 2] |  |  | -0.78 (0.15)*** |
| ROA[quartile 3] |  |  | -1.87 (0.24)*** |
| ROA[quartile 4] |  |  | -1.71 (0.22)*** |
| Debt2Equity[quartile 2] |  |  | -0.12 (0.17) |
| Debt2Equity[quartile 3] |  |  | -0.84 (0.20)*** |
| Debt2Equity[quartile 4] |  |  | -0.40 (0.18)** |
| YearOfEntry[2014] | 0.11 (0.21) | 0.11 (0.21) | 0.10 (0.21) |
| YearOfEntry[2015] | 0.73 (0.23)*** | 0.73 (0.23)*** | 0.61 (0.23)*** |
| YearOfEntry[2015] | 0.86 (0.26)*** | 0.86 (0.26)*** | 0.67 (0.28)** |
| YearOfEntry[2015] | 1.49 (0.31)*** |  |  |
| YearOfEntry[2015] | 1.06 (0.80) |  |  |
| Num. obs. | 9453 | 5463 | 5010 |
| Missings | 5 | 5 | 458 |
| Num. events | 294 | 260 | 241 |
| AIC | 4695.90 | 4079.72 | 3598.62 |
| Num. events | 294 | 260 | 241 |
| PH test | 0.90 | 0.72 | 0.72 |

***$p < 0.01$, **$p < 0.05$, *$p < 0.1$

Being a green ISU positively and significantly influences the survival chances in all the three models. In particular, the hazard ratios, computed as an exponential function of the coefficient, reports that green startups have a failure rate equal to 53.2% if compared to the others (Model 3) while controlling for ROA, D/E and sector of activity. This means that a green company is more than twice as likely to survive than a non-green company.

Moreover, focusing on the sector of activity, both Model 1 and Model 2 report significance of coefficients and hazard ratios, indicating a higher probability of leaving the market for services, tourism and trade sectors, compared to manufacturing. The result is not surprising, considering that the startups operating in the manufacturing sector are those with the most survival chances.

Although the results are in line with the KM curves, it is necessary to highlight a loss of statistical significance of the coefficients introducing the control variables. As ROA increases, the instantaneous exit potential decreases. The same consideration may be done for D/E, which seems to have a strongly negative impact on the survival performance of innovative startups.

Model 3 has the smallest value for the AIC criterion and hence has to be preferred because it has a relatively better fit. Finally, the Grambsch-Therneau test (P.H. test) has been also performed to assess the hypothesis of proportional hazards assumption. For all three models the P.H. test statics is not significant, with an associated *p*-value greater than 0.10, and hence the proportional hazards assumption is respected.

## 5. Discussion

In last twenty years, Western economies have experienced a deep cycle of stagnation. Italy is one of the countries most affected by all these issues. Startups may represent a possible key to open the gates of recovery, in particular the most innovative ones and those that dedicate the right attention to the environment. However, newborn companies may experience a premature failure from the market. Hence, the implementation of strong forms of protection to support startups seems to be an unavoidable necessity. The present paper aimed at investigating if Italian regulations to boost the birth of startups have been successful and if these newborn companies show survival rates different from other businesses. Special attention has been dedicated to green startups, in order to understand their differences with non-green ones. Data and implemented analyses show that Italian startups turning their production to greenness tend to survive longer than their counterparts, leading to conclude that attention to the environment and rewards more newborn companies.